\documentclass[pra,twocolumn,
superscriptaddress,
showpacs,
aps
]{revtex4-1}
\usepackage{graphicx}
\usepackage{amsmath}
\usepackage{amsfonts}
\usepackage{amssymb}
\usepackage{hyperref}
\usepackage{color}
\usepackage{bbold}

\newcommand{\beq}{\begin{equation}}
\newcommand{\eeq}{\end{equation}}

\usepackage{dsfont}
\usepackage{placeins} 
\usepackage{bm}           

\begin{document}
	
\title{Edge states of the long-range Kitaev chain: an analytical study}
	\author{Simon B. J\"ager} 
   \affiliation{Theoretische Physik, Universit\"at des Saarlandes, D-66123 Saarbr\"ucken, Germany} 
   \affiliation{JILA and Department of Physics, University of Colorado, Boulder, Colorado 80309-0440, USA.}
   \author{Luca Dell'Anna}
\affiliation{Dipartimento di Fisica e Astronomia G. Galilei, Universit\`a degli studi di Padova, via Marzolo 8, 35131 Padova, Italy}
	\author{Giovanna Morigi} 
	\affiliation{Theoretische Physik, Universit\"at des Saarlandes, D-66123 Saarbr\"ucken, Germany}
\begin{abstract}
We analyze the properties of the edge states of the one-dimensional Kitaev model with long-range anisotropic pairing and tunneling. Tunneling and pairing are assumed to decay algebraically with exponents $\alpha$ and $\beta$, respectively, and $\alpha,\beta>1$. We determine analytically the decay of the edges modes. We show that the decay is exponential for $\alpha=\beta$ and when the coefficients scaling tunneling and pairing terms are equal. Otherwise, the decay is exponential at sufficiently short distances and then algebraic at the asymptotics. We show that the exponent of the algebraic tail is determined by the smallest exponent between $\alpha$ and $\beta$. Our predictions are in agreement with numerical results found by exact diagonalization and in the literature.
\end{abstract}
	\maketitle
\section{Introduction}

Topological phases of matter have been attracting great interest since the discovery of the integer and fractional quantum Hall effect \cite{Klitzing:1980, Fractional:Hall:Effect} and of topological phase transitions \cite{NobelpriceTop:2016}. Topological superconductors \cite{Hasan:2010,Bernegiv:2013}, in particular, offer promising perspectives for realizing robust quantum devices \cite{Kitaev2008,Nayak2008,Stern:2010} due to the presence of topologically protected states, the so-called edge modes \cite{Kitaev2001}. In this context, the Kitaev chain is a theoretical model which exhibits topological order \cite{Kitaev2001} and can be mapped the Ising model with nearest-neighbor interactions \cite{Greiter:2014}. For open boundaries and in the topological non-trivial phase the Kitaev model supports the existence of zero energy excitations, the Majorana edge modes  \cite{Kitaev2001}. These modes are spatially localized at the chain's edge and are an indicator for the non-trivial topological nature of this phase \cite{Ryu:2002}. 

An extension of the Kitaev model has been recently discussed which describes algebraic decay of the tunneling and/or pairing terms 	\cite{Pientka2013,Klinovaja2013,Pientka2014,Neupert2016,Vodola2014,Vodola2016,Alecce2017}. This model is expected to describe experimental realizations of long-range topological superconductors \cite{Nadj-Perge2014,Pawlak2016,Ruby2017}. It has been shown that the long-range interactions leads to a modification of the phase diagram \cite{Vodola2014,Vodola2016,Alecce2017}. Moreover, numerical studies of this model revealed algebraically localized edge states and an algebraic closing of the energy gap \cite{Vodola2014,Vodola2016}. When the pairing and tunneling terms are isotropic, instead, 
exponential localization is recovered independently of the power law exponent, as long as it is larger than unity \cite{Vodola2016}. 

In this paper we perform an analytical study of the spatial localization of the Majorana edge states in the long-range Kitaev models. Our analysis includes anisotropic and isotropic pairing and tunneling terms with the same or different algebraic exponents. For this model, we determine analytically the asymptotic scaling of the edge modes' tails. Our findings are supported by numerical calculations that are in good agreement with the numerical results reported in the literature for specific parameter choices \cite{Vodola2014,Vodola2016,Alecce2017,Viyuela2015,Viyuela:2018fpv}.

This paper is structured as follows. In Sec.~\ref{sec:1} we introduce the long-range Kitaev model and the Majorana operators. Here, we discuss the formalism which is the starting point of our analysis. In Sec.~\ref{sec:2} the basic equation determining the edge states is derived, which allows us to determine their site occupation inside the bulk. We determine a general expression that gives the behavior of the zero eigenmodes away from the edges as a function of the parameters of the model. In section~\ref{sec:3} we compare our analytical findings with numerical results and in Sec. \ref{sec:conclusions} we draw the conclusions.

\section{Long-range Kitaev chain}
\label{sec:1}

We consider $N$ polarized Fermions on a lattice with open boundary conditions. Their Hamiltonian has the form of a Kitaev model with long-range interactions: 
\begin{align}
\hat{H}=-\sum_{n=1}^N\left[\sum_{r=1}^{N-n}\left(j_r^{\alpha}\hat{c}_n^{\dag}\hat{c}_{n+r}+\Delta_r^{\beta}\hat{c}_n^{\dag}\hat{c}_{n+r}^{\dag}+\mathrm{H.c.}\right)+\mu\hat{c}_n^{\dag}\hat{c}_n\right]\,,\label{LRK}
\end{align}
where $\hat{c}_i$ and $\hat{c}_i^{\dag}$ are the Fermionic annihilation and creation operators that fulfill the anticommutation relations $\{\hat{c}_i,\hat{c}_j\}=\{\hat{c}_i^{\dag},\hat{c}_j^{\dag}\}=0$, ${\{\hat{c}_i^{\dag},\hat{c}_j\}=\delta_{i,j}}$, and $\delta_{i,j}$ is the Kronecker delta. Here, $\mu$ is the chemical potential. The other coefficients $j_r^{\alpha}$ and $\Delta_r^{\beta}$ scale the tunneling and pairing terms between two sites at distance $r$. They depend on the distance according to the power law decay given by: 
\begin{eqnarray}
&&j_r^{\alpha}=\frac{J}{N_\alpha}\frac{1}{r^{\alpha}},\\
&&\Delta_r^{\beta}=\frac{\Delta}{N_\beta}\frac{1}{r^{\beta}}\,,
\end{eqnarray}
with the exponent $\alpha,\beta>1$. We denote the parameters $J$ and $\Delta$ by tunneling rate and the pairing strength, respectively. The coefficient $N_{\gamma}=\sum_{r=1}^{N}r^{-\gamma}$ warrants normalization.  For $\gamma=\alpha,\beta>1$, which is the case we consider, $N_{\gamma}\to\zeta(\gamma)$ for $N\to\infty$ and $\zeta(\gamma)$ is the Riemann zeta function \cite{Olver:2010}. 

For sufficiently fast decaying interaction and hopping terms the system possesses two different phases separated by the quantum critical point $\mu_c=2J$ \cite{Kitaev2001}.  In the thermodynamic limit the two topological phases can be distinguished by the bulk topological invariant $w$: For $|\mu|>\mu_{c}$ the ground state is non-degenerate and $w=0$; in the nontrivial phase $|\mu|<\mu_{c}$ the bulk topological invariant is $w=1$,  the ground state is doubly degenerate, and can support Majorana edge modes. At finite size $N$ the spectrum is always gapped. In this work we analyze the spatial localization of the Majorana edge modes for a chain with open boundaries and as a function of the exponents $\alpha$ and $\beta$. 

\subsection{Bogoliubov-de Gennes Hamiltonian}

The Hamiltonian in Eq.~\eqref{LRK} is quadratic and can be cast into a compact form by introducing the  $2N$ component vector operator  $\hat{\bf C}=(\hat c_1,\hat c_1^\dagger,\ldots,\hat c_N, \hat c_N^\dagger)^T$: 
\begin{align}
\label{H:C}
\hat{H}=\hat{\bf C}^{\dag}\hat{H}_{\mathrm{BdG}}\hat{\bf C}\,
\end{align}
where $\hat H_{\mathrm{BdG}}$ is the  Bogoliubov-de Gennes Hamiltonian, and is a $2N\times 2N$ matrix. In order to give its form in a compact way, first we write the vector-operator $\hat{\bf C}$ as
\begin{align}
\label{C:n}
\hat{\bf C}=\sum_{n=1}^N\left[|1,n\rangle\hat{c}_n+|0,n\rangle\hat{c}_n^{\dag}\right],
\end{align}
where ${|1,n\rangle\hat{c}_n+|0,n\rangle\hat{c}_n^{\dag}=(0,\ldots,0,\hat c_n,\hat c_n^\dagger,0,\ldots,0)^T}$ is a $2N$ component vector whose elements are all zeroes except for the $2n$ and the $2n+1$ components. Using this definition, the Bogoliubov-de Gennes Hamiltonian can be written as
\begin{align}
\label{H:BdG}
\hat{H}_{\mathrm{BdG}}=-\mu\,\hat{\tau}_z\otimes \mathbb{1}_N-\hat{\tau}_z\otimes \hat{J}^{\alpha}-\hat{\tau}_y\otimes \hat{\Delta}^{\beta}\,.
\end{align}
In Eq.~\eqref{H:BdG} we have introduced the $N\times N$ matrices
\begin{eqnarray}
&&\mathbb{1}_N=\sum_{n=1}^N|n\rangle\langle n|\,,\\
&&\hat{J}^\alpha=\sum_{n=1}^{N} \sum_{r=1}^{N-n}j^{\alpha}_r\left(|n\rangle\langle n+r|+|n+r\rangle\langle n|\right)\,,\label{J}\\
&&\hat{\Delta}^\beta=\sum_{n=1}^{N} \sum_{r=1}^{N-n}i\Delta^{\beta}_r\left(|n\rangle\langle n+r|-|n+r\rangle\langle n|\right)\label{D}\,,
\end{eqnarray}
and the Pauli matrices
\begin{align*}
\hat{\tau}_x=&\frac{|1\rangle\langle 0|+|0\rangle\langle 1|}{2}\,,\\
\hat{\tau}_y=&\frac{|1\rangle\langle 0|-|0\rangle\langle 1|}{2i}\,,\\
\hat{\tau}_z=&\frac{|1\rangle\langle 1|-|0\rangle\langle 0|}{2}\,,
\end{align*}
with $[\hat{\tau}_l,\hat{\tau}_m]=i\sum_{n}\epsilon_{lmn}\tau_n$, where $\epsilon_{lmn}$ is the Levi-Civita tensor and the indices are $n,m,l\in\{x,y,z\}$.

\subsection{Majorana Fermions}

In this work we are interested in the properties of Majorana edge states. Majorana edge states are eigenstates of Hamiltonian~\eqref{LRK}, their eigenvalue vanishes in the thermodynamic limit $N\to\infty$. This property becomes evident when the Hamiltonian in presented in terms of the Majorana operators. The Majorana operators are hermitian Fermionic operators, are here denoted by the operators $\hat{\gamma}_j$ and $\hat{\gamma}_{j+N}$ ($j=1,2,\dots N)$ and are defined by the relations
\begin{align*}
&\hat{\gamma}_j=\frac{\hat{c}_j+\hat{c}_{j}^\dagger}{\sqrt{2}},\\
&\hat{\gamma}_{N+j}=i\frac{\hat{c}_j^\dagger-\hat{c}_{j}}{\sqrt{2}}\,,\\
\end{align*}
with $\{\hat{\gamma}_i,\hat{\gamma}_j\}=\delta_{i,j}$. For convenience, we define the Majorana vector operator
\begin{align}
\label{def:gamma}
\hat{\boldsymbol{\gamma}}=\sum_{n=1}^N\left[|1,n\rangle\hat{\gamma}_n+|0,n\rangle\hat{\gamma}_{n+N}\right].
\end{align}
The Majorana vector operator is connected to $\hat{\bf C}$ by the relation 
\begin{align}
\label{C:gamma}
\hat{\bf C}=&\frac{1}{\sqrt{2}}\left[\begin{pmatrix}
1 &i\\
1& -i
\end{pmatrix}\otimes\mathbb{1}_N\right]\hat{\boldsymbol{\gamma}}\equiv A\hat\gamma\,,
\end{align}
where the last equality defines the matrix $A=A_2\otimes \mathbb{1}_N$ connecting the two representations, with
\begin{align}
A_2=\frac{1}{\sqrt{2}}\begin{pmatrix}
1 &i\\
1& -i
\end{pmatrix}\,.
\end{align}
Using Eq.~\eqref{C:gamma} we rewrite the Hamiltonian, Eq.~\eqref{H:C}, as
\begin{equation}
\hat{H}=\hat{\boldsymbol{\gamma}}^{\dag}\hat{H}_{M}\hat{\boldsymbol{\gamma}}\,,
\end{equation}
where
\begin{align}
\hat{H}_{M}&=A^\dagger H_{\rm BdG}A\nonumber\\
&=\mu\hat{\tau}_y\otimes\mathbb{1}_N+\hat{\tau}_y\otimes \hat{J}^{\alpha}+\hat{\tau}_x\otimes \hat{\Delta}^{\beta}\,.\label{H:M}
\end{align}
To obtain this expression we have used that $A_2^\dagger\hat{\tau}_xA_2=\hat\tau_z$, $A_2^\dagger\hat{\tau}_yA_2=-\hat\tau_x$, and $A_2^\dagger\hat{\tau}_zA_2=-\hat\tau_y$.

For later convenience we further elaborate on this notation. We  define the generalized lowering operators by
\begin{align}
\hat{J}^{\alpha}_{-}=&\sum_{n=1}^N\sum_{r=1}^{N-n}j_r^{\alpha}|n\rangle\langle n+r|,\\
\hat{\Delta}^{\beta}_{-}=&\sum_{n=1}^N\sum_{r=1}^{N-n}\Delta_r^{\beta}|n\rangle\langle n+r|\,.
\end{align}
The raising operators are the hermitian conjugate: ${\hat{J}_{+}^{\alpha}=(\hat{J}_{-}^{\alpha})^{\dag}}$, ${\hat{\Delta}_{+}^{\beta}=(\hat{\Delta}_{-}^{\beta})^{\dag}}$. Using these definitions, we rewrite Eq.~\eqref{J} and Eq.~\eqref{D} as
\begin{align}
\hat{J}^{\alpha}=\hat{J}_{+}^{\alpha}+\hat{J}_{-}^{\alpha},
\end{align}
and
\begin{align}
\hat{\Delta}^{\beta}=-i\hat{\Delta}_{+}^{\beta}+i\hat{\Delta}_{-}^{\beta},
\end{align}
respectively. The Bogoliubov-de Gennes Hamiltonian for the Majorana vector operators takes then the form
\begin{align}
\hat{H}_{M}=\mu\,\hat{\tau}_y\otimes\mathbb{1}_N&+i\left(\hat{\tau}_x\otimes \hat{\Delta}^{\beta}_{-}-i\hat{\tau}_y\otimes \hat{J}_{-}^{\alpha}\right)\nonumber\\
&-i\left(\hat{\tau}_x\otimes \hat{\Delta}^{\beta}_{+}+i\hat{\tau}_y\otimes \hat{J}_{+}^{\alpha}\right)\,.\label{HM}
\end{align}
This is the Hamiltonian form at the basis of our treatment. 

\subsection{Majorana edge states in the isotropic Kitaev chain with vanishing chemical potential}

We now consider first the case of the isotropic Kitaev chain with  $\mu=0$. We thus set $\Delta=J$ in Eq.~\eqref{HM} and take $\alpha=\beta$, namely, tunneling and pairing decay with the same power-law exponent. The  corresponding Hamiltonian reads
\begin{align}
\hat{H}_0&=i\left(\hat{\tau}_-\otimes \hat{J}^{\alpha}_{-}\right)-i\left(\hat{\tau}_+\otimes \hat{J}^{\alpha}_{+}\right),\label{H0}
\end{align}
where $\hat{\tau}_{-}=\hat{\tau}_x-i\hat{\tau}_y$ and $\hat{\tau}_{+}=\hat{\tau}_-^{\dag}$ are the spin lowering and raising operators, respectively. The Majorana edge states are the eigenstates of $\hat{H}_0$ with zero eigenenergy: 
\begin{align}
|e_1\rangle=&|1,1\rangle,\\
|e_2\rangle=&|0,N\rangle\,.
\end{align}
They are the exact eigenstates for any power-law exponent $\alpha$. Independently of the exponent, they are localized at edge of the Kitaev chain, at site $n=1$ or $n=N$, respectively. From definition \eqref{def:gamma} they correspond to the Majorana modes $\hat{\gamma}_1$ and $\hat{\gamma}_{2N}$. These operators  commute with the Hamiltonian $\hat{\boldsymbol{\gamma}}^{\dag}\hat{H}_{0}\hat{\boldsymbol{\gamma}}$. These edge states still exist for finite but small $\mu$ and even for the anisotropic Kitaev chains. The analysis of the properties of the edge states in the anisotropic Kitaev chain is the aim of the next section.

\section{Edge modes in the anisotropic Kitaev chain}
\label{sec:2}

In this section we discuss the properties of the edge modes when the Kitaev chain is anisotropic, namely, when the exponent of the power law decay are not equal, $\alpha\neq \beta$, and/or the coefficients of tunneling and pairing differ, $J\neq \Delta$. We thus seek for eigenmodes $\hat\gamma_E$ that commute with $\hat{H}$, Eq.~\eqref{LRK}, in the thermodynamic limit $N\to\infty$:
\begin{align}
\label{Eq:Edge}
\left[\hat{H},\hat\gamma_E\right]=0\,.
\end{align}
We write the eigenmodes as a linear superposition of the modes $\hat\gamma_j$ with scalar coefficients $h_j$:
\begin{equation}
\hat\gamma_E=\sum_{\ell=1}^{2N}h_\ell\,\hat\gamma_\ell={\bf h}^T\hat{\boldsymbol{\gamma}}=\hat{\boldsymbol{\gamma}}^T{\bf h}\,,
\end{equation}
where we have introduced the vector ${{\bf h}=(h_1,h_2,...,h_{2N})^T}$. Edge modes are characterized by coeffcients $h_\ell$ whose modulus is expected to be maximum close to the edges, which are here given by $\ell=1$ and $\ell=2N$, and which shall decrease with the distance from the closer edge. From Eq.~\eqref{Eq:Edge} we can identify some general requirements that these coefficients shall fulfill. Since $\hat{H}$ is quadratic the commutator on the left-hand-side of Eq.~\eqref{Eq:Edge} can be rewritten as
\begin{align}
\left[\hat{H},{\bf h}^T\hat{\boldsymbol{\gamma}}\right]=-\left[{\bf h}^T\hat{H}_{M}\hat{\boldsymbol{\gamma}}-\hat{\boldsymbol{\gamma}}^T\hat{H}_{M}{\bf h}\right]=2\hat{\boldsymbol{\gamma}}^T\hat{H}_{M}{\bf h}\,,
\end{align}
where in the last equality we have used  that $\hat{H}_{M}^T=-\hat{H}_{M}$. Therefore, Eq.~\eqref{Eq:Edge} takes the form of the eigenvalue equation
\begin{align}
\label{Eq:Edge:h}
\hat{H}_{M}{\bf h}=0\,.
\end{align}
For the remainder of this section we want to present how we intend to derive ${\bf h}$. 

\subsection{Eigenvalue equation}

Let us first formulate the eigenvalue problem
\begin{align}
\label{Eq:Edge:h:2}
\hat{H}_{M}|v\rangle=E|v\rangle\,,
\end{align}
where we use Dirac's notation of operator for matrices and ket (bra) vectors for right (left) eigenvectors. We define the Hamiltonian
\begin{align}
\hat{H}_1=\hat{H}_M-\hat{H}_0\,,
\end{align}
such that $\hat{H}_M=\hat{H}_0+\hat{H}_1$.  For $\hat{H}_1=0$ eigenstates of $\hat{H}_{M}$ at $E=0$ are the vectors $|v_0\rangle =a|e_1\rangle+b|e_2\rangle$, with $a$ and $b$ scalars. In particular, vectors $|e_1\rangle$ and $|e_2\rangle$ form a basis for the kernel of $\hat{H}_0$ for $J\neq0$. 

We now consider the perturbation and search for eigenstates with $E\approx 0$, which are localized at the edges. For this purpose we introduce the projector $\hat{\mathcal P}$ into the subspace spanned by $|e_1\rangle$ and $|e_2\rangle$, and the projector $\hat{\mathcal Q}=\mathbb{1}-\hat{\mathcal P}$ onto the orthogonal complement of the kernel of $\hat{H}_0$, such that $\hat{\mathcal Q}+\hat{\mathcal P}={\mathbb{1}}$ is the unity matrix $\mathbb{1}$. We note that $\hat{\mathcal P}$ projects on the edges, while $\hat{\mathcal Q}$ contains the chain's bulk.

We now make some general considerations. By taking the projection of Eq.~\eqref{Eq:Edge:h:2} on the subspace corresponding to $\hat{\mathcal P}$ and using that $\hat{\mathcal P}\hat H_0=\hat H_0\hat{\mathcal P}=0$ we obtain the relation 
\begin{align}
\hat{\mathcal P}(\hat H_1-E)|v\rangle=0\,,\label{generalizedEigenvalue}
\end{align} 
We note that, since we expect $E\approx 0$, then the scalar product $|\langle v_0|\hat H_1|v\rangle|\ll 1$, thus the states we search for have small overlap with the bulk. We further find the structure of $|v\rangle$ by considering that Eq.~\eqref{Eq:Edge:h:2} can be rewritten as
\begin{align*}
 E|v\rangle=&\hat H_M|v\rangle\\
=&\left[\hat{\mathcal Q}\hat H_0+\hat{\mathcal Q}(\hat{H}_1-E \mathbb{1})\right]|v\rangle+[\hat{\mathcal Q}E+\hat{\mathcal P}\hat{H}_1]|v\rangle\\
=&\hat{\mathcal Q}\hat H_0\left[\mathbb{1}+(\hat{\mathcal Q}\hat{H}_0)^{-1}\hat{\mathcal Q}(\hat{H}_1-E\mathbb{1})\right]|v\rangle\\
&+[E+\hat{\mathcal P}(\hat{H}_1-E\mathbb{1})]|v\rangle
\,.
\end{align*}
Using Eq. \eqref{generalizedEigenvalue} we can write the eigenstates as a function of $|v_0\rangle$ such that
\begin{align}
|v\rangle=\frac{1}{\mathbb{1}+(\hat{\mathcal{Q}}\hat{H}_0)^{-1}\hat{\mathcal{Q}}(\hat{H}_1-E\mathbb{1})}|v_0\rangle\,,\label{v}
\end{align}
which ensures that $\big[\mathbb{1}+(\hat{\mathcal Q}\hat{H}_0)^{-1}\hat{\mathcal Q}(\hat{H}_1-E\mathbb{1})\big]|v\rangle=|v_0\rangle$.

Substituting Eq. \eqref{v} in Eq. \eqref{generalizedEigenvalue}, we can now cast the problem into finding the kernel of operator $\hat \Gamma$:
\begin{align}
\hat \Gamma=\hat{\mathcal{P}}(\hat{H}_1-E)\frac{1}{\mathbb{1}+(\hat{\mathcal{Q}}\hat{H}_0)^{-1}\hat{\mathcal{Q}}(\hat{H}_1-E\mathbb{1})}\hat{\mathcal{P}}\,,\label{Gamma}
\end{align}
namely, we shall find the eigenvalues $E$ for which $\hat \Gamma$ has a non-trivial kernel. In particular, for edge state we should obtain $E=0$ in the thermodynamic limit $N\to\infty$.

Let us now use the explicit forms of $\hat{H}_0$ and $\hat{H}_1$ to calculate the matrix ${\big[\mathbb{1}+(\hat{\mathcal{Q}}\hat{H}_0)^{-1}\hat{\mathcal{Q}}(H_1-E\mathbb{1})\big]^{-1}}$. This can be performed after considering that $\hat{\mathcal{Q}}$ projects on the bulk of the chain. 
Therefore, in the thermodynamic limit we can apply the Fourier transform to derive the spectrum and the eigenvectors of $\hat{H}_M$ and $\hat{H}_1$. For this purpose we define the $k$ vectors
\begin{align*}
|k\rangle=\frac{1}{\sqrt{N}}\sum_{n=1}^Ne^{i(n-1)k}|n\rangle,
\end{align*}
with $k=2\pi m/N$ and $m=0,1,\dots,N-1$. Then, 
\begin{align*}
\hat{\mathcal{Q}}\hat{H}_{0}=\sum_{k}\begin{pmatrix}
0&\frac{-iJ}{\zeta(\alpha)}\mathrm{Li}_\alpha(e^{-ik})\\
\frac{iJ}{\zeta(\alpha)}\mathrm{Li}_\alpha(e^{ik})&0
\end{pmatrix}\otimes|k\rangle\langle k|
\end{align*}
and 
\begin{align*}
\hat{\mathcal{Q}}\hat{H}_{M}=\sum_{k}\begin{pmatrix}
0&-iF(-k)\\
iF(k)&0
\end{pmatrix}\otimes|k\rangle\langle k|\,,
\end{align*}
with
\begin{align}
F(k)=\frac{\mu}{2}&+\frac{\Delta}{2\zeta(\beta)}\left[\mathrm{Li}_\beta(e^{ik})-\mathrm{Li}_\beta(e^{-ik})\right]\nonumber\\
&+\frac{J}{2\zeta(\alpha)}\left[\mathrm{Li}_\alpha(e^{ik})+\mathrm{Li}_\alpha(e^{-ik})\right]\,.\label{Fk}
\end{align}
Here, $\mathrm{Li}_{\gamma}(z)=\sum_{r=1}^\infty z^r/r^\gamma$ denotes the Polylogarithm and $|z|\leq1$ \cite{Olver:2010}. 

Now the problem of finding the inverse in Eq.~\eqref{Gamma} reduces to invert $2\times 2$ matrices. This can be solved analytically and gives, for $E=0$, 
\begin{align}
\frac{1}{\mathbb{1}+(\hat{\mathcal{Q}}\hat{H}_0)^{-1}\mathcal{Q}H_1}=&\sum_k\begin{pmatrix}
{\mathcal M}(k)&0\\
0&{\mathcal M}(-k)
\end{pmatrix}\otimes |k\rangle\langle k|\,,\label{M}
\end{align}
with 
\begin{align}
{\mathcal M}(k)=\frac{J}{\zeta(\alpha)}\,\frac{\mathrm{Li}_\alpha(e^{ik})}{F(k)}\,.\label{Mk}
\end{align}
Later on we will check the consistency of the assumption that the eigenvectors have vanishing energy by estimating the scaling of $E$ with the chain's size.

\subsection{Localization of the edge modes along the chain}

\begin{figure}[h!]
	\center \includegraphics[width=0.70\linewidth]{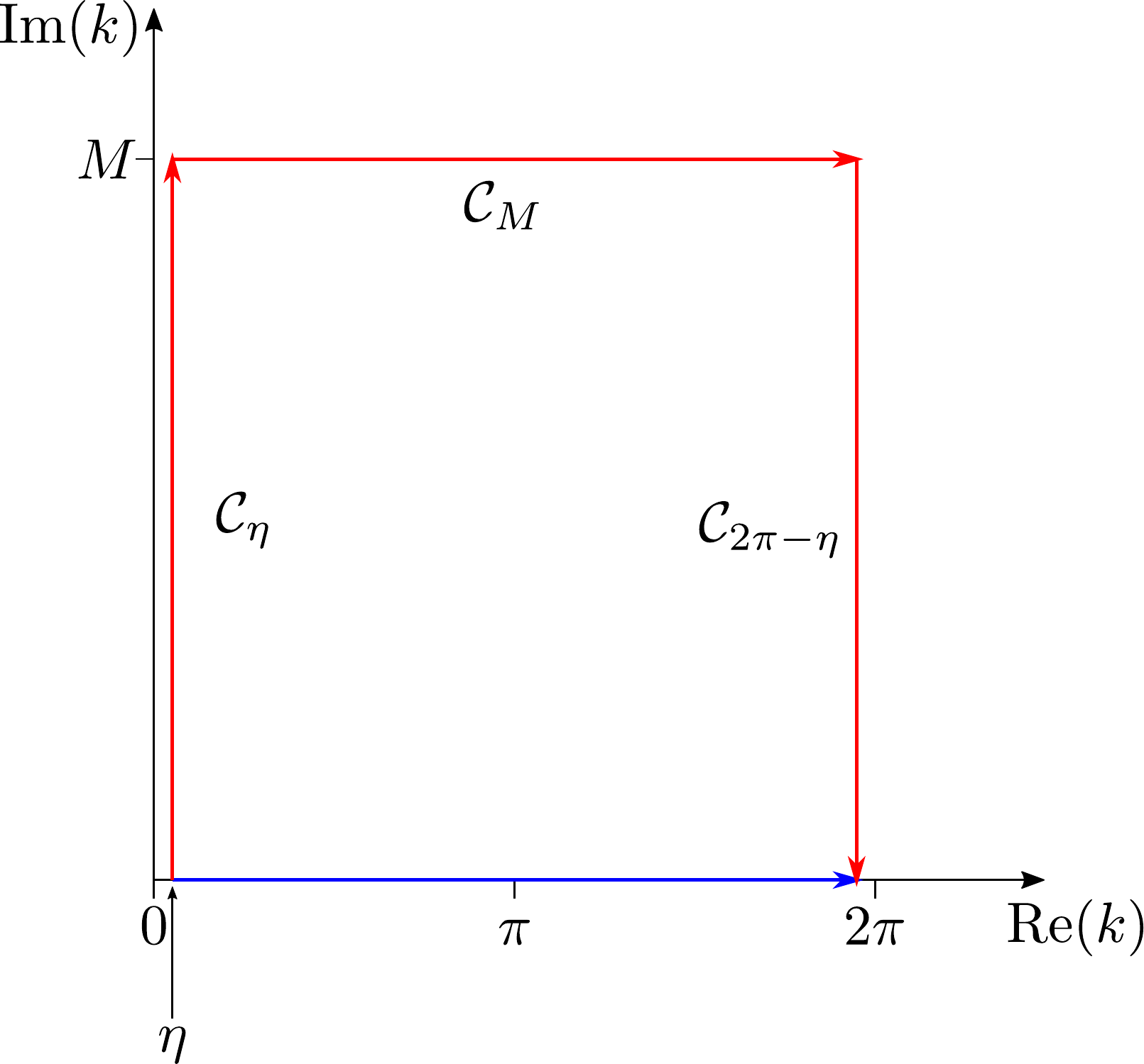}
	\caption{Sketch of the contour integration \eqref{In}. Here, the blue line indicates the integration path $\mathcal{C}(y)=y$ along the real axis ($y\in[\eta,2\pi-\eta]$), the red line shows the contour. Here, $\mathcal{C}_{\eta}(y)=\eta+iy$ with $y\in[0,M]$, $\mathcal{C}_{M}(y)=y+iM$ with $y\in[\eta,2\pi-\eta]$, and $\mathcal{C}_{2\pi-\eta}(y)=2\pi-\eta+i(M-y)$ with $y\in[0,M]$. \label{Fig:contour}}
\end{figure}

We can now determine the probability amplitudes $h_{\ell}$ for the eigenstates at zero eigenvalue. This reduces to calculate the projection of the edge state that we will call $|e\rangle$, which is $|v\rangle$ given by Eq.~\eqref{v} at $E=0$, onto the bulk position $|n\rangle\equiv |1,n\rangle$. We take $|v_0\rangle=|e_1\rangle\equiv |1,1\rangle$ and calculate the amplitude
\begin{align}
\langle n| e\rangle= A(n)
\end{align}
where
\begin{align}
\label{Cn}
A(n)=\frac{1}{2\pi}\int_{0}^{2\pi}dk\,e^{i(n-1)k}{\mathcal M}(k)\,.
\end{align}
The function $|A(n)|^2$ gives the spatial occupation of the chain site $n$ of the eigenmode at zero eigenvalue. By construction, for   $\hat{H}_M=\hat{H}_0$ it reduces to the edge state $|e_1\rangle$, namely, $A(n)=\delta_{n,1}$. As it has been used here, the probability amplitude $A(n)$ can be found taking the continuum limit, when the summation $\sum_{k}$ goes over to a continuum of $k$-values in the interval $[0,2\pi)$. Using the residue theorem it can be rewritten as the sum of the integral along the contour, $I(n)$, and of the residues, $R(n)$, as it follows
\begin{align}
A(n)= R(n)+I(n).
\label{Cnequ}
\end{align}
The integral is taken along the contour illustrated in Fig.~\ref{Fig:contour} and reads
\begin{align}
I(n)=\lim_{\eta\rightarrow 0^+}\left[\int_{\mathcal{C}_\eta}+ \int_{\mathcal{C}_{2\pi-\eta}}+ \int_{\mathcal{C}_M}\right] dk\,\frac{e^{i(n-1)k}}{2\pi}{\mathcal M}(k)\,,\label{In}
\end{align}
where the interval of integration along the imaginary axis and the real axis are $[0,M]$ ($M>0$) and $[\eta,2\pi-\eta]$, respectively. The individual paths are $\mathcal{C}_{\eta}(y)=\eta+iy$ with $y\in[0,M]$, $\mathcal{C}_{M}(y)=y+iM$ with $y\in[\eta,2\pi-\eta]$, and $\mathcal{C}_{2\pi-\eta}(y)=2\pi-\eta+i(M-y)$ with $y\in[0,M]$. 
The summation over the residues in Eq.~\eqref{Cnequ} goes over the complex numbers $k_0$ with ${0<\mathrm{Re}(k_0)<2\pi}$ and ${\mathrm{Im}(k_0)>0}$ for which $\mathrm{Res}[e^{i(n-1)k}{\mathcal M}(k),k_0]$ does not vanish and reads
\begin{align}
R(n)=i\hspace{-0.15cm}\sum_{k_0:\mathrm{Im}(k_0)>0}\mathrm{Res}\big[e^{i(n-1)k}{\mathcal M}(k),k_0\big]\,.\label{Rn}
\end{align}
Both residues and integral contributions determine the behavior of $|e\rangle$ at the chain bulk. Below we argue that the residues contributes to $A(n)$ with an exponential decay (see Eq.~\eqref{R:exp}), while the integral term is different from zero only in the anisotropic Kitaev chain. In this case its contribution is an algebraic decay with the smallest exponent, either $\alpha$ or $\beta$ (see Eq.~\eqref{algebraic}). 

\subsubsection{Residues}

By inspection of Eq.~\eqref{Mk} we observe that the only values where $\mathrm{Res}[e^{i(n-1)k}{\mathcal M}(k),k_0]$ does not vanish are the zeros of $F(k)$, Eq.~\eqref{Fk}. Therefore we search for the values $k_0$ such that  $F(k_0)=0$. Within the area of the contour ${\mathcal M}(k)$ is a meromorphic function and we can take its Laurent series about the root $k_0$ of $F(k)$:
\begin{align}
{\mathcal M}(k)=\sum_{\ell=-\infty}^{\infty}a_{\ell}(k_0)(k-k_0)^{\ell}\,,\label{Laurent}
\end{align}
where the $a_{\ell}$ are the coefficients. Moreover, since  ${\mathcal M}(k)$ is meromorphic there exists a finite and positive index $L_{k_0}$ such that $a_{\ell}(k_0)=0$ for $\ell<-L_{k_0}$. This index determines  the order of the pole of ${\mathcal M}(k)$ at $k_0$.

Using Eq.~\eqref{Laurent} the residues at $k_0$ can be expressed as 
\begin{align}
\label{R:fig}
\mathrm{Res}\big[e^{i(n-1)k}{\mathcal M}(k),k_0\big]=e^{i(n-1)k_0}P_{k_0}(n)\,,
\end{align}
where $P_{k_0}(n)$ is a polynomial in $n$ which depends on the coefficients of the Laurent expansion as
\begin{align}
P_{k_0}(n)=\sum_{\ell=0}^{L_{k_0}-1}\frac{[i(n-1)]^{\ell}}{\ell!}a_{-1-\ell}(k_0),
\end{align}
using the expansion of $e^{i(n-1)(k-k_0)}$ close to $k_0$. 
Since all poles are isolated, then the behavior of Eq.~\eqref{Rn} in the bulk is dominated by the residue at the point $k_0'$ with the smallest imaginary part, namely ${\rm Im}(k_0')=\xi$ is such that 
${\xi\le {\rm Im}(k_0)}$ for all $k_0$. We distinguish two cases, when $\xi>0$ and when instead $\xi=0$. For $\xi>0$  then for $n\gg 1$ the sum over the residues Eq.~\eqref{Rn} behaves as
\begin{align}
\label{R:exp}
R(n) \sim e^{-\xi n}\tilde P_{k_0'}(n)\,,
\end{align}
%
where $\tilde P_{k_0'}(n)= iP_{k_0'}(n)e^{i(n-1){\rm Re}(k_0')}$.  If $k_0'$ is a pole of order one, then the polynomial $P_{k_0'}(n)$ is simply a constant independent on $n$.

When $\xi=0$, a pole lies on the real axis. This occurs at the critical point, where there is no localized mode. This treatment allows us to determine the critical values.
To derive the phase diagram it is sufficient to find the zeros of $F(k)$ in Eq.~\eqref{Fk} which lie on the real axis in the interval $k\in[0,2\pi]$. The gap closes at $k=0$ and $k=\pi$. For $k=\pi$ we obtain the threshold
\begin{align}
\mu=2J\left(1-2^{1-\alpha}\right).
\end{align}
For $k=0$ we obtain 
\begin{align}
\mu=-2J,
\end{align}
therefore the topological phase is bounded as follows
\begin{align}
-2J\leq\mu\leq 2J\left(1-2^{1-\alpha}\right),
\label{Boundaries}
\end{align}
in agreement with Ref.~\cite{Alecce2017}. Note that the boundaries are independent of the power-law exponent of 
the pairing term, and, for $\alpha\to\infty$, one recovers the standard topological phase, in agreement with the 
findings in Ref.~\cite{Vodola2014,Vodola2016,Alecce2017}.
When $\mu$ is inside this interval, the residues contributes with a function which exponentially decays away from the edges. 
For any $\eta>0$, $k=0$ is excluded, therefore a good candidate for $k_0^\prime$, the pole which contributes more to the sum of residues, can be searched above $k=\pi$.

\subsubsection{Integrals}
We will now extract the behavior of the integrals in Eq.~\eqref{In}. For this purpose we use that $\int_{\mathcal{C}_M}dke^{i(n-1)k}{\mathcal M}(k)$ vanishes in the limit $M\to\infty$. In this limit 
the integral to solve is 
\begin{align}
I(n)&=-\frac{1}{\pi}\int_{0}^\infty e^{-y(n-1)}{\rm Im}({\mathcal M}(iy))\,.\label{Integral}
\end{align} 
Here, ${\mathcal M}(k)$ is given in Eq.~\eqref{Mk}, and its imaginary part specifically reads:
\begin{align}
\label{Mk:Im}
\mathrm{Im}({\mathcal M}(iy))=-\frac{J}{\zeta(\alpha)}\, {\rm Im}(F(iy))\,\frac{\mathrm{Li}_\alpha(e^{-y})}{|F(iy)|^2}\,,
\end{align}
with the function $F(k)$ of Eq.~\eqref{Fk}.


In order to determine the behavior for $n\gg 1$, we expand $\mathrm{Im}({\mathcal M}(iy))$ in leading order of $y$ using the Taylor expansion of the Polylogarithm \cite{Olver:2010}:
\begin{align*}
\mathrm{Li}_\gamma(e^{-y})=&\Gamma(1-\gamma)y^{\gamma-1}+\sum_{k=0}^\infty\frac{\zeta(\gamma-k)}{k!}(-y)^k,\\
\mathrm{Li}_\gamma(e^{y})=&\Gamma(1-\gamma)\cos(\pi(\gamma-1))y^{\gamma-1}+\sum_{k=0}^\infty\frac{\zeta(\gamma-k)}{k!}y^k\nonumber\\&+i\Gamma(1-\gamma)\sin(\pi(\gamma-1))y^{\gamma-1}\,.
\end{align*}
where the real part is here only well-defined for $\gamma\notin \mathbb{N}$, while the coefficient of the imaginary part is ${\Gamma(1-\gamma)\sin(\pi(1-\gamma))}={\pi/\Gamma(\gamma)}$. 
In leading order in the expansion, Eq. \eqref{Mk:Im} is given by 
\begin{align*}
\mathrm{Im}({\mathcal M}(iy))\approx\Lambda\left[\frac{J}{2\zeta(\alpha)\Gamma(\alpha)}y^{\alpha-1}-\frac{\Delta}{2\zeta(\beta)\Gamma(\beta)}y^{\beta-1}\right]\,,
\end{align*}
and $\Lambda=J\pi/(\mu/2+J)^2$.
Substituting in Eq. \eqref{Integral} we obtain
\begin{align}
I(n)&\approx-\left[\frac{J}{2\zeta(\alpha)}n^{-\alpha}-\frac{\Delta}{2\zeta(\beta)}n^{-\beta}\right]\frac{J}{\left(\frac{\mu}{2}+J\right)^2}\,,\label{algebraic}
\end{align}
which is valid for $n\gg 1$. This expression shows that the integral vanishes for  $J=\Delta$ and $\alpha=\beta$. In this case the decay of the edge state is dominated by the residues and is purely exponential. 
Otherwise, when $J\neq\Delta$ or $\alpha\neq\beta$ the behavior of $I(n)$ is dominated by an algebraic decay with the smallest exponent between $\alpha$ and $\beta$. In this case the contribution of the integral determines the decay of the edge mode in the bulk. This result is in agreement with the behavior reported in the specific limits \cite{Vodola2014,Vodola2016,Alecce2017}.

\subsection{Discussion}

We have analyzed the properties of the spatial dependence of the zero eigenmodes focusing on their behavior away from the edges. Our study shows that their form is determined by the properties of the integral \eqref{Cn}, whose integrand is determined by the ratio between the coefficients of $\hat H_0$ and of $\hat H_M$ in Fourier space. In particular, the asymptotic scaling is determined by the properties of the coefficients of the Hamiltonian $\hat H_M$, while the properties of the coefficients of $\hat{H}_0$ enter into the scaling factors. 

The results we obtained did not make use of perturbation theory. Nevertheless they were derived under the assumption that the resulting eigenvector $|v\rangle$ are at zero energy, so that we could set $E=0$  in Eq.~\eqref{v}, that we called $|e\rangle$. We now verify that the result we obtain is consistent with this assumption. For this purpose we show that the eigenvector $|e\rangle$ we have found fulfills Eq.~\eqref{generalizedEigenvalue} at $E=0$. We first determine the scalar product $\langle v_0'|\hat{H}_1|e\rangle$ where $|v_0^\prime\rangle=a |e_1\rangle+b  |e_2\rangle$ and 
\begin{align}
\hat{H}_1|e\rangle=\sum_{k}i\left[F(k)-\frac{J}{\zeta(\alpha)}\mathrm{Li}_{\alpha}(e^{-ik})\right]{\mathcal M}(k)\begin{pmatrix}
0\\
1
\end{pmatrix}\otimes|k\rangle\,,
\end{align} 
where we took $|v_0\rangle=|e_1\rangle$ in Eq.~\eqref{v}. For $|v_0^\prime\rangle= |e_2\rangle$ the overlap is maximal and reads 
\begin{align}
&\langle e_2|\hat{H}_1|e\rangle\nonumber\\
&=\int_{0}^{2\pi}dk\frac{ie^{i(N-1)k}}{2\pi}\frac{\left[F(k)-\frac{J}{\zeta(\alpha)}\mathrm{Li}_{\alpha}(e^{ik})\right]\frac{J}{\zeta(\alpha)}\mathrm{Li}_{\alpha}(e^{ik})}{F(k)}.\label{Energyestimate}
\end{align} 
This integral can be calculated using the residues theorem as above. It is composed by the sum of an exponential scaling $\exp(-\xi N)$ and an algebraic scaling $N^{-\mathrm{min}\{\alpha,\beta\}}$, where the second contribution vanishes for  $\alpha=\beta$ and $\Delta=J$. This provides an estimate how the energy scales for every finite $N$ value and shows that the assumption $E=0$ is correct in the thermodynamic limit.

\section{Numerical analysis of the edge modes}
\label{sec:3}

In this section we investigate numerically the spatial distribution of the edge states and their energy. This is done by determining the smallest positive eigenenergy $E_{\mathrm{edge}}$ of $\hat{H}_{M}$, Eq.~\eqref{HM}, and the corresponding edge state $|e\rangle$, in a chain of $N=8000$ sites. We analyze in particular the probability 
\[
\mathcal P_n=|A(n)|^2=|\langle n|e\rangle|^2
\]
that the eigenstate $|e\rangle$ occupies site $n$.

\begin{figure}[ht]
	\center \includegraphics[width=1\linewidth]{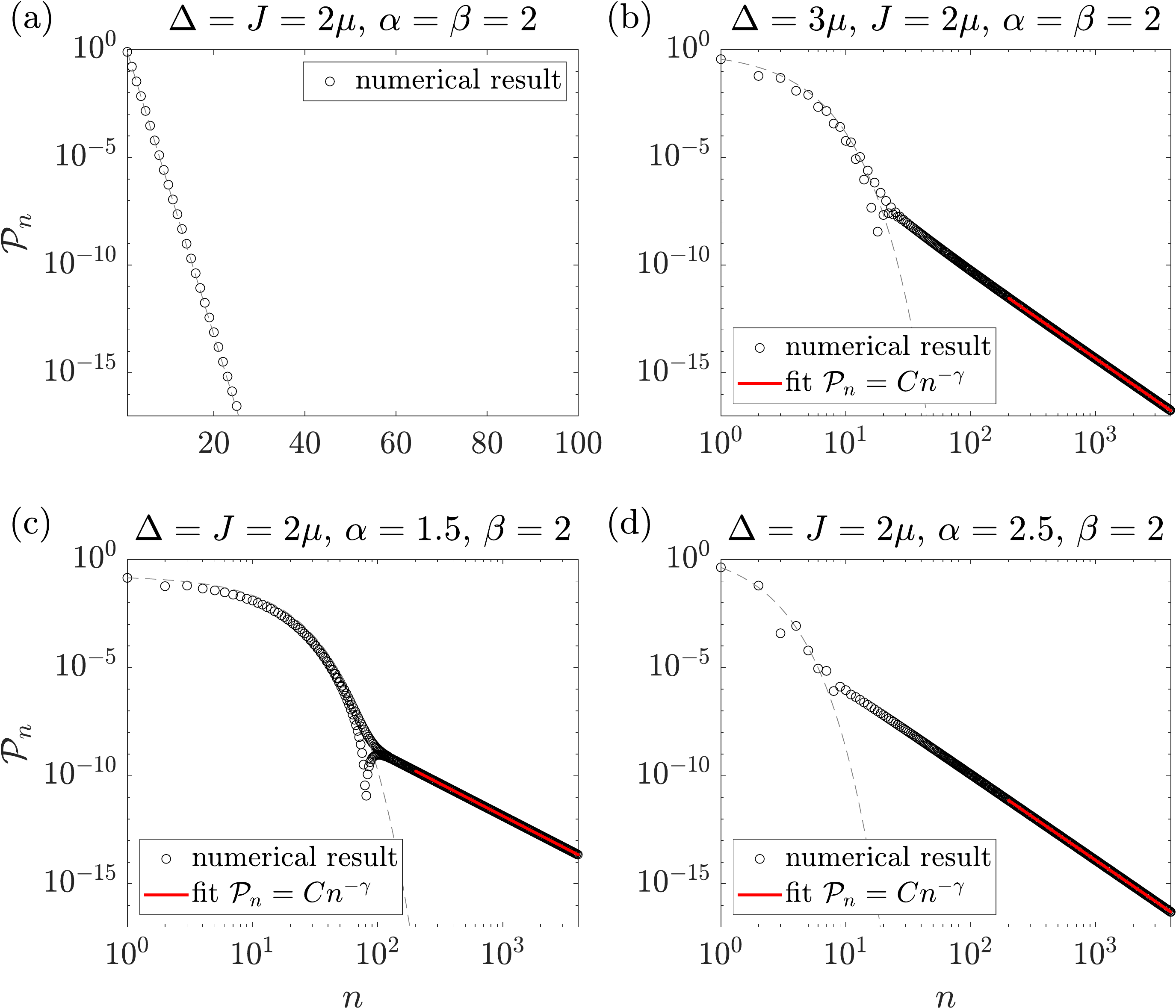}
	\caption{Probability that the edge state occupies the site $n$, $\mathcal P_n=|\langle n|e\rangle|^2$, as a function of $n$ and for different choices of the parameters. The parameters are given on top of each subplot. The results have been evaluated by numerically diagonalizing the matrix $\hat H_M$, Eq.~\eqref{HM}.  The gray dashed lines show the exponential behavior $\mathcal{P}_n=\mathcal{P}_1e^{-2\xi (n-1)}$. The exponent $\xi$ is  found by numerically determining  the pole $k_0'$ of ${\mathcal M}(k)$ with the smallest imaginary part. The tails of the curves, in the anisotropic cases, fitted by the function $\mathcal{P}_n=Cn^{-\gamma}$, are shown by the red line in subplots (b)-(d), the interval of the fit is from $n=200$ to $n=4000$. We find $\gamma=4.03$~(b), $\gamma=2.98$~(c), and $\gamma=3.98$~(d), with a rounding error to two digits after the decimal point.\label{Fig:tails}}
\end{figure}

Figure~\ref{Fig:tails} displays the probability $\mathcal P_n$ as a function of $n$ and for different values of $\Delta,J$ and $\alpha,\beta$. The behavior of subplot (a) corresponds to a parameter choice with $\Delta=J$ and $\alpha=\beta$ and displays an exponential decay of the edge mode. Subplots (b)-(d) correspond to different choices of parameters for the anisotropic Kitaev chain, where the algebraic decay at large $n$ is visible. We have fitted the numerical results at the asymptotics using the fitting function 
\[
\mathcal P_n\simeq C \,n^{-\gamma}
\] 
finding $\gamma$ in good agreement with the analytical result
$\gamma=2\,\mathrm{min}\{\alpha,\beta\}$. 
The effect of the contribution $R(n)$ of the residues in Eq.~\eqref{Cnequ} is also visible in Fig.~\ref{Fig:tails}. In order to show that, we find the pole $k_0'$ of ${\mathcal M}(k)$ with the smallest imaginary part, calculate $\xi=i\textrm{Im}(k_0')$, and plot 
\[
\mathcal {P}_n\simeq \mathcal{P}_1 e^{-2 \xi (n-1)} 
\]
as gray dashed line. While we observe for the isotropic case (subplot (a)) very good agreement in the full range of $n$, we see that the exponential behavior is still in good agreement for small $n$ in the anisotropic case (subplots (b)-(d)). However, for larger values of $n$, in the anisotropic case, we observe eventually that the algebraic tail becomes dominant as expected.

\begin{figure}[ht]
	\center \includegraphics[width=1\linewidth]{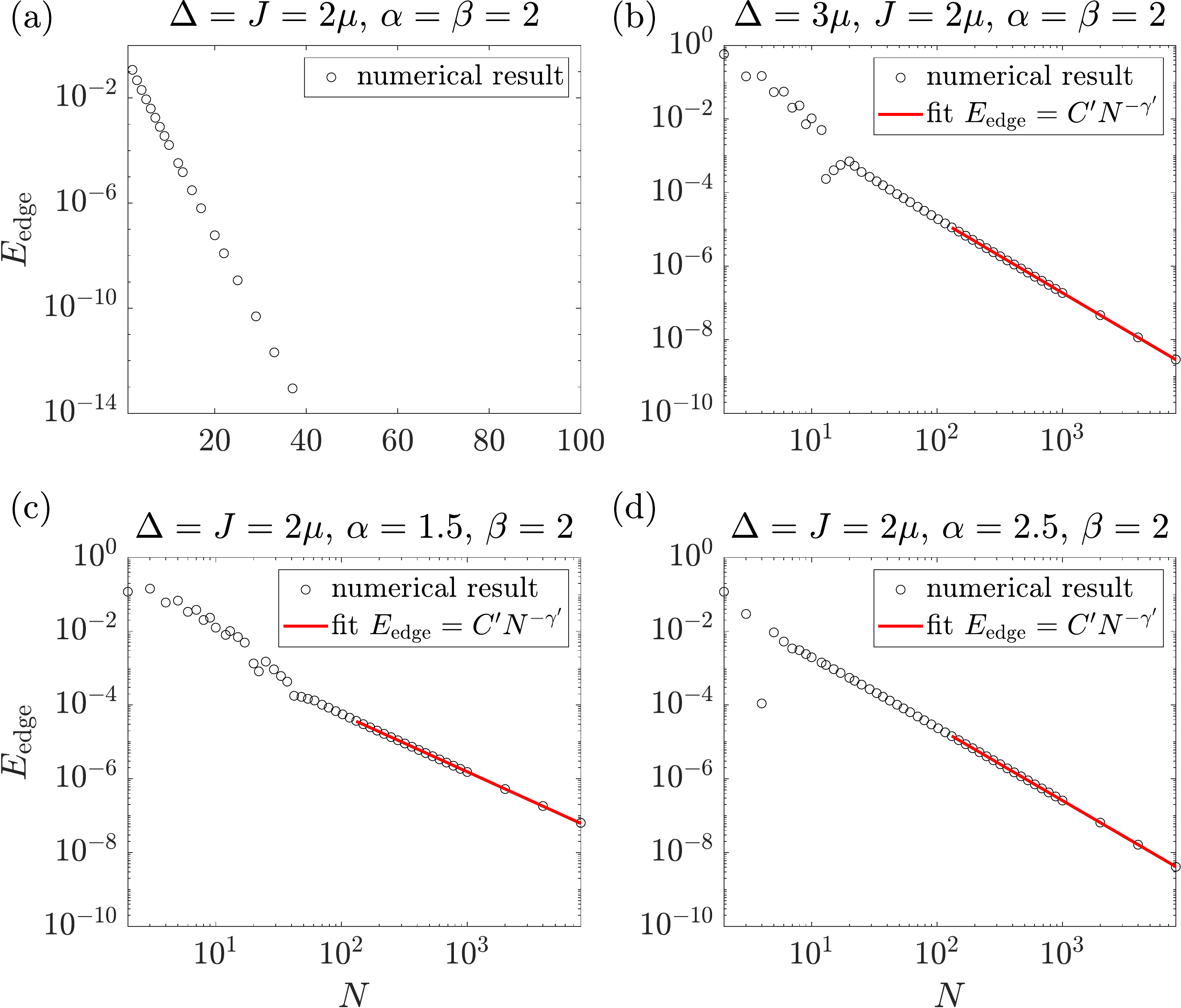}
	\caption{The energy $E_{\mathrm{edge}}$ of the edge state as a function of the system size $N$ and for different choices of the parameters. The parameters are given on top of each subplot. The results have been evaluated by numerically diagonalizing the matrix $\hat H_M$, Eq.~\eqref{HM}. The parts of the curves fitted by the function $C^{\prime}N^{-\gamma^{\prime}}$ are shown by the red line in subplots (b)-(d), the interval of the fit goes from $N=131$ to $N=8000$. We find $\gamma^{\prime}=2.01$~(b), $\gamma^{\prime}=1.55$~(c), and $\gamma^{\prime}=1.98$~(d), with a rounding error to two digits after the decimal point.\label{Fig:energy}}
\end{figure}

Figure~\ref{Fig:energy} displays the scaling of the energy of the eigenmode $|e\rangle$ with the chain size $N$ and for different values of $\Delta,J$, and $\alpha,\beta$. Subplot~\ref{Fig:energy}(a) is calculated for the same parameters as subplot~\ref{Fig:tails}(a) but different sizes and displays an exponential scaling. In the anisotropic Kitaev chain, subplots (b)-(d), the scaling is algebraic, the fitting function at large $N$ gives 
\begin{align}
E_{\mathrm{edge}} \sim C^{\prime}N^{-\gamma^{\prime}}\label{Fitenergy}
\end{align}
with $\gamma^{\prime}\approx\mathrm{min}\{\alpha,\beta\}$ in good agreement with our analytical result.

\section{Conclusions}
\label{sec:conclusions}

We have analytically determined the spatial decay of the edge states in the long-range Kitaev chain. The expressions we obtain are general and hold for any choice of the tunneling and pairing rates and of the exponents of the algebraic decay, as long as these are larger than unity. Our result allows to determine the boundaries of the topological non-trivial phase. Within this phase, it predicts that the edge modes are exponentially localized at the chain edges in the isotropic case, namely, when the pairing and tunneling rates are equal and decay with the same exponent. By means of this model we can extrapolate the characteristic length as a function of the parameters. This behavior agrees with the numerical results reported in Refs.~\cite{Vodola2014,Vodola2016}. Algebraic decay of the edge modes is instead found in the anisotropic case, when either the exponent and/or the rates of tunneling and pairing differ. In this case, the smallest exponent determines the algebraic scaling of the tails, while at shorted distances the decay is exponential. This behavior has been observed for some specific cases ~\cite{Vodola2014,Vodola2016,Alecce2017}. Our result is analytical and generalizes these findings.  

Our approach could be generalized to higher dimensional cases, e.g. two or three dimensional systems \cite{Zhang:2019}. It could be extended in order to describe the out-of-equilibrium behavior following slow quenches across the critical point Ref.~\cite{Defenu:2019}. 
\acknowledgements
The authors acknowledge stimulating discussions with Michael Kaicher, Nicol\'o Defenu, and Tilman Enss. This work has been supported by the German Research Foundation (the priority program No. 1929 GiRyd), by the European  Commission (ITN ColOpt) and by the German Ministry of Education and Research (BMBF) via the QuantERA project NAQUAS. Project NAQUAS has
received funding from the QuantERA ERA-NET Cofund in Quantum Technologies implemented within the European Union's Horizon 2020 program. We acknowledge support from the DARPA and ARO grant W911NF-16-1-0576.

\end{document}